# Ultrahigh measured unipolar strain > 2 % in polycrystalline bulk piezoceramics: Effects of disc dimension

Gobinda Das Adhikary and Rajeev Ranjan*

Department of Materials Engineering, Indian Institute of Science, Bangalore-560012, India.

**Abstract**

Bulk polycrystalline piezoelectric ceramics are extensively used in wide ranging applications as actuators, transducers, and sensors. For actuator applications, it is desirable that the piezoelectric ceramic gives as large electric field driven strain as possible. In recent years there is an increasing interest in the design and development of such piezoceramic materials. Here we show that piezoceramic discs (10 mm diameter) of even simple ferroelectric systems like PZT and BaTiO$_3$-based piezoelectric can show very large strain ~ 2 – 3 % application of electric field when the thickness is reduced to ~ 200 microns or below. This is accompanied by increasing asymmetry in the bipolar strain-field loops. The measurements were carried out in the conventional manner using Radiant's ferroelectric testing system and MTI-Fotonic displacement sensor. The motivation of this brief report is make the research community aware of this important effect in piezoceramics and encourage discussion/research on the this interesting phenomenon, which is applicable to all piezoceramics.

Email: rajeev@iisc.ac.in

I. Introduction

Piezoelectric materials change shape on application of electric-field (converse effect) and develop voltage on application of force. Both these aspects are used in a variety of technological applications. Single crystal Quartz is well-known for its piezoelectric property. Most advanced piezoelectrics exhibit much higher electromechanical response than quartz and are most often solid solutions of ferroelectric perovskites such as Pb(Zr$_x$Ti$_{1-x}$)O$_3$ (commonly abbreviated as PZT) [1]. Large electromechanical response in such solid solutions occur near specific composition generally known as morphotropic phase boundary (MPB) composition [2]. For PZT, the MPB is at x=0.52 [2]. The enhancement in the electromechanical response at the MPB is attributed to the ease of polarization rotation and switching of the ferroelectric-ferroelastic domains [3, 4]. For actuator applications, it is desirable that the piezoceramic exhibit as large electric field driven unipolar strain as possible in repeated cycles. Large unipolar strain (~ 1.7 %) has been reported in single crystals of relaxor ferroelectric based systems like Pb(Zn$_{1/3}$Nb$_{2/3}$)O$_3$-PbTiO$_3$ [5]. Most polycrystalline piezoceramics, including PZT, are reported to show unipolar strain in the range 0.2 – 0.4 % [6]. Unipolar strain greater than 0.4 % in non-textured polycrystalline piezoceramic was first reported by Zhang et al in $Na_{0.5}Bi_{0.5}TiO_3$ (NBT)-based lead-free piezoceramic [7]. A higher value of 0.7 % was later reported by Liu and Tan in a differently modified NBT system [8]. Narayan et al reported unipolar strain of ~1 % in a La-modified $BiFeO_3$-$PbTiO_3$ system [9]. While there is a tendency to attribute the large electrostrain to the special composition of the material, the mechanisms (such are reversible switching of ferroelastic domains, field induced phase change [10], field induced ergodic to ferroelectric transition [9]) are not specific to those compositions. Similar mechanisms can be found in several other solid solutions for which no large electric field driven strain have been reported. In this brief report we show that the measured electrostrain is highly sensitive to the thickness of the disc shaped specimen used in the measurements.

## 2. Experimental

Here we demonstrate the effect on three piezoelectric material systems $Pb(Zr_xTi_{1-x})O_3$, $Ba(Ti_{1-y}Sn_y)O_3$ and $(1-z)Na_{0.5}Bi_{0.5}TiO_3$-$(z)SrTiO_3$. Dense polycrystalline ceramics of these systems were made using the conventional solid state synthesis method. The ceramics were ~ 95 % of the theoretical density. All sintered ceramic discs used in these measurements were 10 mm diameter. The as sintered discs were ~ 1mm thick. The electrostrain measurements were performed by gradually reducing the thickness of the discs by rubbing the surfaces with fine emery paper by almost equal amount (from both sides). Before performing the electrostrain measurements, the pellets were thermally annealed (~ 500 C) to remove the residual stresses accumulated during the polishing process. The surfaces of the pellets were electrode by applying silver paint. Ferroelectric measurements were carried out using the Radiant Premier Precision II ferroelectric testing system. Field driven strain measurements were carried out with a MTI Fotonic displacement sensor and sample holder assembly supplied by Radiant, Inc. The upper and bottom electrodes of the sample holder was ~ 6 mm diameter. During the measurements, the center of the circular disc shaped pellets (10 mm diameter) was kept almost at the center of the electrodes of the sample holder. For ferroelectric and strain measurements, triangular wave form of 1 Hz was applied.

## 3. Results and Discussion

Fig. 1 shows the thickness dependence bipolar strain, unipolar strain and polarization-electric field hysteresis loops of PZT x=0.51. This composition corresponds to the tetragonal border of the MPB region of PZT [2]. The pellet of 0.7 mm (700 micron) thickness exhibits unipolar electrostrain of 0.2 % at 50 kV/cm. The unipolar strain increases to 0.9 % when the thickness was reduced to 300 micron. At 200 micron the unipolar increases to ~ 1.5 % at 50 kV/cm. When the amplitude of the field was increased to 70 kV/cm, the measured unipolar strain on the 200 micron disc reached 2.5 % (last row of figures in Fig. 1). The corresponding effective converse piezoelectric coefficient, defined as $d_{33}^* = S_{max}/E_{max}$ is ~ 3600 pm/V. An interesting point to note is that the bipolar strain developed asymmetry as the strain increases on decreasing the pellet thickness. Some variations in the shape of the polarization-electric field hysteresis loop with decreasing thickness -the polarization shows increasing tendency for saturation with decreasing thickness, Fig 1c.

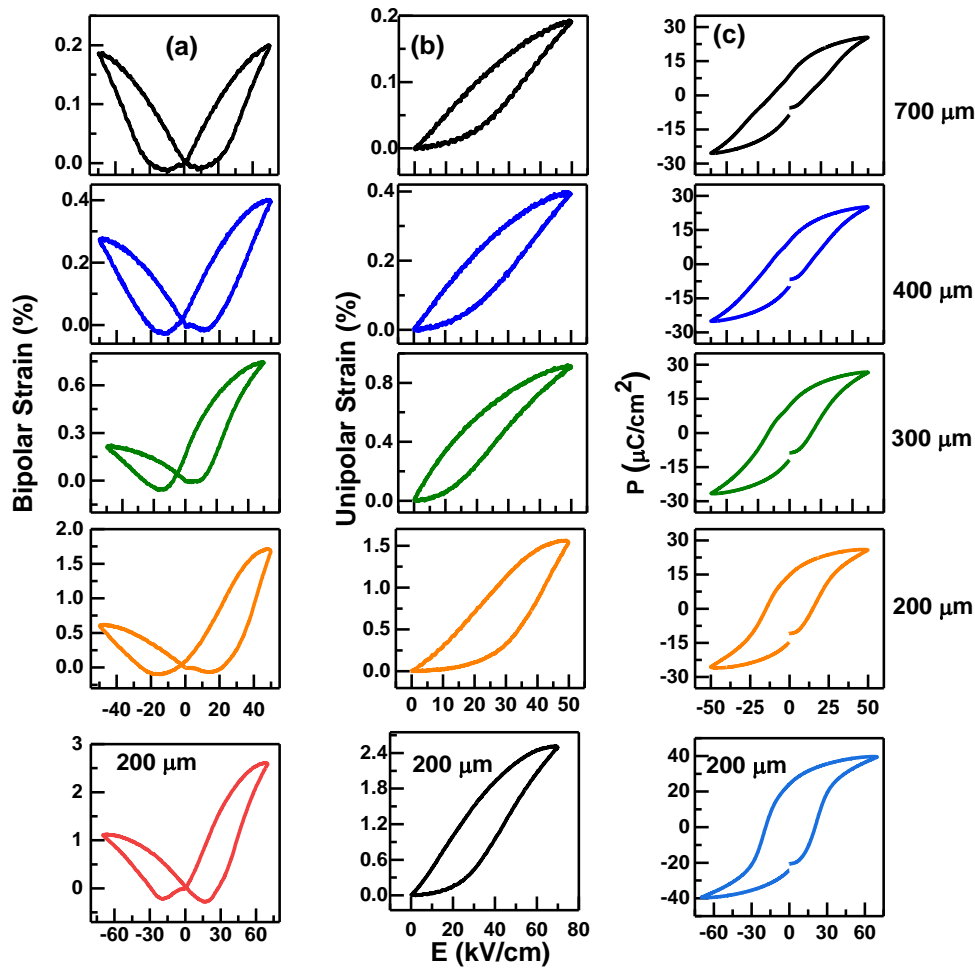

Fig. 1: (a) bipolar strain for different thickness of PZT: x=0.51. (b) unipolar strain of different thickness (c) Polarization-electric field loops of different thickness of PZT:x=0.51.

Fig. 2 shows the effect of varying pellet thickness on the strain measured on a tetragonal composition (y=0.01) at the border of the tetragonal + orthorhombic phase coexistence region of $BaTi_{1-y}Sn_yO_3$ [11, 12]. As compared to PZT, the coercive field of BTS: y=0.01 is significantly smaller. As a consequence, large increase in the measured electrostrain can be seen even at lower field, say 30 kV/cm. The measured unipolar strain of the 700-micron BTS:y=0.01 disc is 0.2 %. The unipolar strain increases sharply below 400 micron and reaches 0.8 % at 300 micron and almost 1.8 % in the 200-micron disc of BTS: y=0.01. When the amplitude of the field was increased to 50 kV/cm, the unipolar strain increased to 2.2 %. The effective converse piezoelectric coefficient $d_{33}^*$ of BTS: y=0.01 reaches ~6000 *pm/V*.

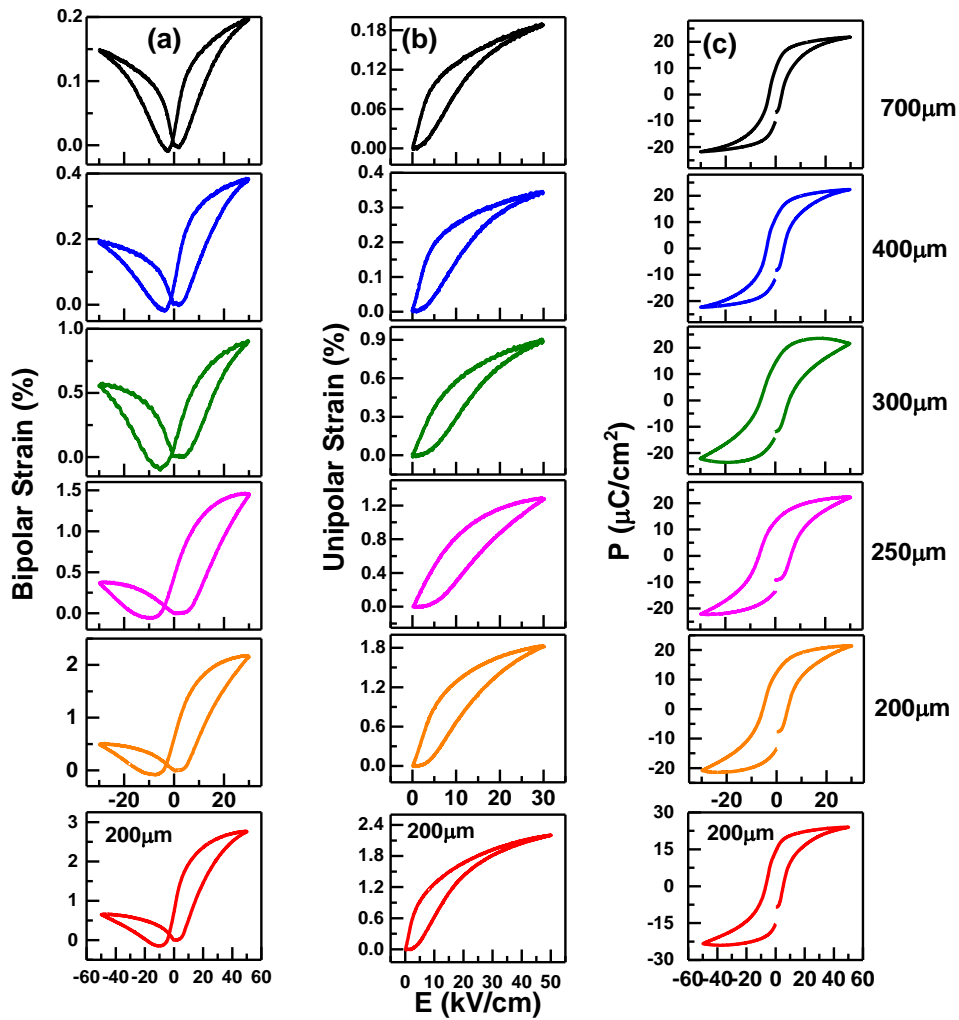

Fig. 2: (a) bipolar strain for different thickness of BaTi$_{0.91}$Sn$_{0.01}$O$_3$. (b) unipolar strain of different thickness of BaTi$_{0.91}$Sn$_{0.01}$O$_3$ (c) Polarization-electric field loops of different thickness of BaTi$_{0.91}$Sn$_{0.01}$O$_3$.

Further, like the PZT composition, the bipolar strain of BTS: y=0.01 develops asymmetry on decreasing thickness, Fig 2a. The area of the P-E loop also increased with decreasing thickness suggesting enhanced domain activity in the thinner discs.

In the last example (Fig. 3), we show the effect of thickness on the measured electrostrain in a Na$_{0.5}$Bi$_{0.5}$TiO$_3$-based polycrystalline piezoceramic, (1-z)Na$_{0.5}$Bi$_{0.5}$TiO$_3$ – (z)SrTiO$_3$, (NBST). Unlike the previous two cases (PZT and BTS), NBST does not exhibit a composition inter-ferroelectric transition since SrTiO$_3$ is a cubic paraelectric. The effect of SrTiO$_3$ is to gradually reduce the rhombohedral ferroelectric distortion and decrease the depolarization temperature.

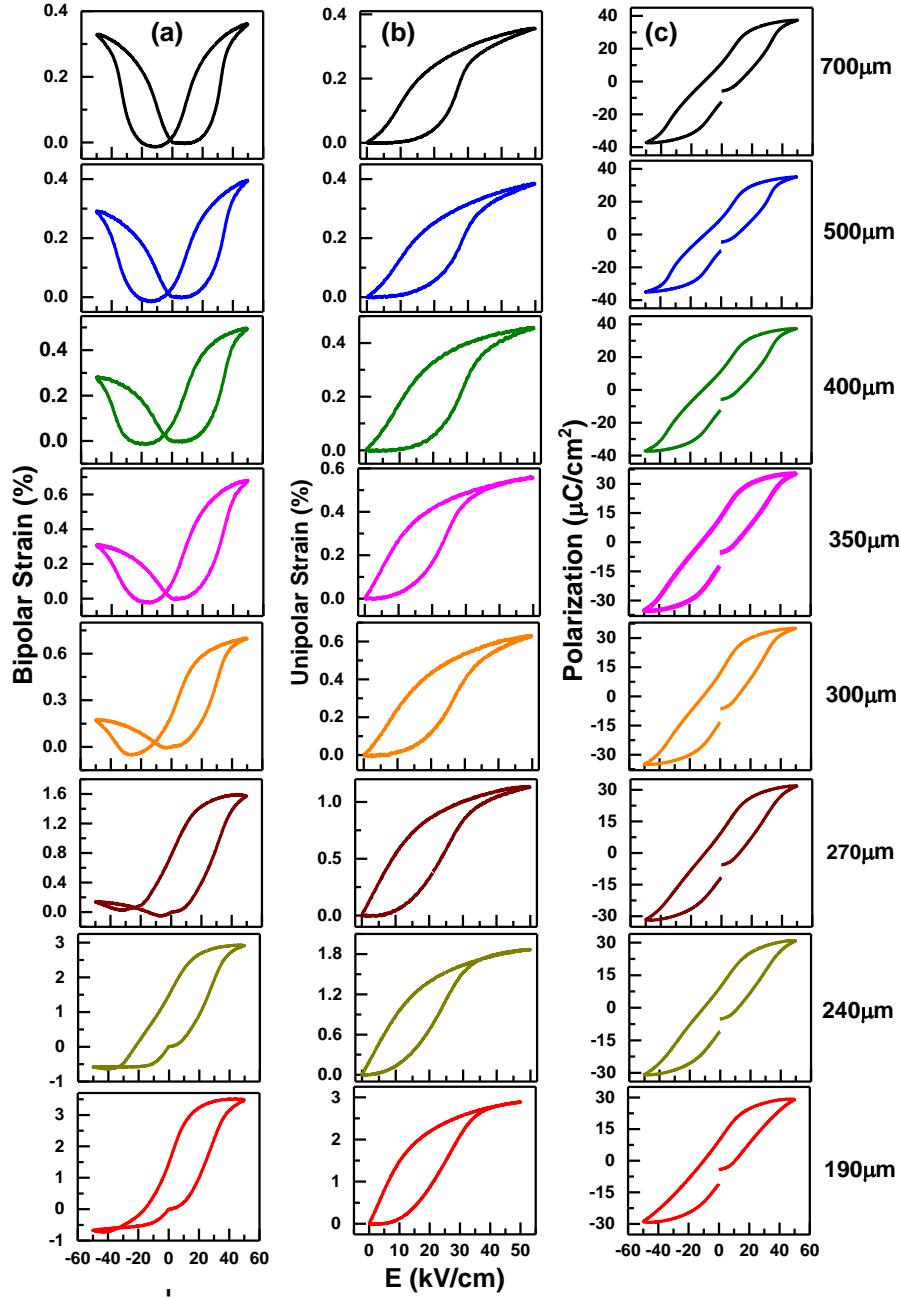

Fig. 3: (a) bipolar strain for different thickness of NBST:z=0.26. (b) unipolar strain of different thickness of NBST:z=0.26 (c) Polarization-electric field loops of different thickness of NBST:z=0.26.

We took a composition NBST: z=0.26. This composition is close to ergodic nonergodic boundary. Fig. 3b shows the unipolar strain of NBST:z=0.26 measured at different thickness. For the 700 micron disc the unipolar strain at 50 kV/cm was measured as 0.35 %. Compared to PZT and BTS (for the same thickness and field amplitude), this value is almost one and half times larger. The measured strain increased phenomenally to 3 % at 50 kV/cm in the 190 micron disc. The corresponding $d_{33}^*$ ~ 6000 pm/V. However, unlike the previous two cases, the asymmetry becomes so acute that the strain becomes negative in the negative cycle (for example for the 240 micron and 190 micron discs, Fig 3a), and the remanent strain shoots up.

Important to note that the shape of the bipolar loops for the 300-micron disc is almost similar to what has been reported in ref [8] for NBT-based system with unipolar strain ~ 0.7 %. Such asymmetric bipolar strain loops are also reported in fatigued and aged ferroelectric specimens [13, 14]. However, the measurements done here were carried out on fresh specimens and not aged/cycled many times.

## 3. Summary and outlook

From the three examples we presented above, it is evident that the electrostrain measured on disc shaped specimen (diameter 10 mm) is highly sensitive to the thickness of the disc. The measured electrostrain reaches unprecedentedly large values 2 – 3 % and the bipolar strain develops asymmetry. In some compositions (not shown here), we have obtained strain ~ 4 - 6 %. Although the maximum strain achieved may vary from specimen to specimen, we would like to emphasize that this effect is common to all piezoceramics we have investigated. In view of these results, it is important to highlight/emphasize the dimension of the specimen while reporting large electrostrain values in piezoelectric ceramics. Although the emphasis here is to present the experimental data as such for the community to appreciate the role of specimen dimension in the measured electrostrain values, there is a need to investigate the possible reasons for the extraordinary strain levels observed in thin bulk piezoceramic discs. Based on the above, we advise caution in attributing large strain measured in polycrystalline piezoceramics to exclusive compositions. It is likely that the dimension of the specimen might be playing a role!

**Acknowledgements:** RR acknowledges the fundings received Science and Engineering Research Board (SERB), Indian Institute of Science (IISc), Naval Research Board (NRB) India, for conducting research on piezoceramics over the years.

**References**

[1] K. Uchino, *Piezoelectric Actuators and Ultrasonic Motors* (Kluwer Academic, Boston, 1996).

[2] B. Jaffe, W. R. Cook, and H. Jaffe, *Piezoelectric Ceramics* (Academic Press, 1971)

[3] D. Damjanovic, *J. Am. Ceram. Soc*. **88**, 2663 (2005).

[4] H. Fu and R. E. Cohen, *Nature* **403**, 281 (2000).

[5] S. -E. Park, T. R. Shrout, *J. Appl. Phys.* **82**, 1804–1811 (1997).

[6] J. Hao, W. Li, J. Zhai, H. Chen, *Mat. Sci. & Engg.* R 135, 1-57 (2019).

[7] S -T. Zhang, A. B. Kounga, E. Aulbach, H. Ehrenberg, J. Rödel, Appl. Phys. Lett. **91**, 112906 (2007).

[8] X. Liu, X. Tan, *Adv. Mater.* **28**, 574 (2016).

[9] B. Narayan, J. S. Malhotra, R. Pandey, K. Yaddanapudi, P. Nukala, B. Dkhil, A. Senyshyn, and R. Ranjan, *Nat. Mater*. **17**, 427 (2018).

[10] M. Hinterstein, M. Hoelzel, J. Rouquette, J. Haines, J. Glaum, H. Kungl, M. Hoffman, Acta Mater., **94**, 319-327 (2015).

[11] A. K Kalyani., K Brajesh., A Senyshyn., R. Ranjan., *Appl. Phys. Lett.* **104**, 252906 (2014)

[12] A. K Kalyani, H. Krishnan, A. Sen, A. Senyshyn, R. Ranjan. *Phys. Rev. B* **91**, 024101 (2015).

[13] J. Nuffer, D. C. Lupascu, J. Rödel, Acta Mater. 48, 3783 (2000)

[14] G. Du, R. Liang, L. Wang, K. Li, W. Zhang, G. Wang, X. Dong, Appl Phys. Lett 102, 162907 (2013)